\documentclass{ws-procs9x6}            

\newcommand{\dvb}{\bar{D}^*}
\newcommand{\jpsi}{J/\psi}

\newcommand{\bma}{\left(\begin{matrix}}
\newcommand{\ema}{\end{matrix}\right)}

\begin{document}
\title{Exploratory study of the coupled-channel scattering in finite volume and the $Z_c(3900)$ resonance}

\author{Mei-Zhu Yan, Dan Zhou$^*$ and Zhi-Hui Guo$^\sharp$}

\address{  Department of Physics and Hebei Advanced Thin Films Laboratory, \\
Hebei Normal University,  Shijiazhuang 050024, China\\
$^*$E-mail: danzhou@hebtu.edu.cn\\ 
$^\sharp$E-mail: zhguo@hebtu.edu.cn}

\begin{abstract}
In this note we first introduce the theoretical formalism of the unitarization of the two-body scattering both 
in the infinite and finite volumes. Then we apply this formalism to the study of the $Z_c(3900)$ resonance in the $J/\psi\,\pi$ and $D\bar{D}^{*}$ coupled-channel scattering. The recent lattice finite-volume spectra are confronted with our predictions. 

\end{abstract}


\bodymatter

\section{Introduction}\label{aba:sec1} 

Important progresses on the study of unstable hadrons in the two-body scattering have been made recently in the lattice QCD computation. The rich and precise finite-volume spectra from the lattice QCD simulations provide valuable information of the underlying hadron interactions, which also give complementary inputs for the phenomenological analyses, compared to the experimental measurements. One powerful way to calculate the physical observables from the finite-volume discrete spectra is the celebrated L\"uscher method~\cite{Luscher:1990ux}, which indeed has been used by many lattice analyses, such as in Refs.~\cite{Dudek:2016cru,Moir:2016srx}. The finite-volume unitarized chiral perturbation theory ($\chi$PT) approach was proposed in Ref.~\cite{Doring:2011vk}, which also has been demonstrated to be useful in the description of the lattice energy levels~\cite{Guo:2016zep,Guo:2018tjx,uchptfv}. 
In Refs.~\cite{Guo:2016zep,Guo:2018tjx}, we have successfully applied the unitarization $\chi$PT approach to describe the lattice energy levels of the $\pi\eta, K\bar{K}, \pi\eta'$ and $D\pi, D\eta, D_s\bar{K}$ coupled-channel scattering, respectively. For the $\pi\eta$ scattering,  the $a_0(980)$ resonance properties and the physical $\pi\eta$ phase shifts and inelasticities are determined~\cite{Guo:2016zep}. After the successful fits to the lattice energy levels of the $D\pi, D\eta, D_s\bar{K}$ scattering in Ref.~\cite{Moir:2016srx}, we predict the physical $D\pi$ scattering amplitudes and determine the properties of the $D^*_0(2400)$ resonance~\cite{Guo:2018tjx}. In this note we first recapitulate the theoretical formalism of the finite-volume unitarized $\chi$PT approach. Then we apply the latter approach to the study of the $\jpsi\,\pi$ and $D\dvb$ coupled-channel scattering and confront our results with the recent lattice energy levels in Refs.~\cite{Cheung:2017tnt,Chen:2019iux}.

\section{Unitarized amplitudes in finite volume}

Following the basic idea of the N/D method, the on-shell two-body scattering amplitudes that satisfy the unitarity condition can be written as~\cite{Oller:1998zr} 
\begin{eqnarray}\label{eq.defut}
 T(s)= [1 + N(s) \cdot G(s)]^{-1}\cdot N(s)\,,
\end{eqnarray}
where by construction the function $G(s)$ only contains the right-hand unitarity cut and the function $N(s)$ only incorporates the crossed-channel dynamics, including the possible local contact terms. One way to write the explicit expression of the function $G(s)$ is  
 \begin{eqnarray}\label{eq.defg}
G(s)=i\int\frac{{\rm d}^4k}{(2\pi)^4}
\frac{1}{(k^2-m_{1}^2+i\epsilon)[(p-k)^2-m_{2}^2+i\epsilon ]}\,, 
\end{eqnarray}
with $s=p^2$. Dimensional regularization can be used to calculate the above integral and the explicit result reads~\cite{Oller:1998zr}
\begin{eqnarray}\label{eq.gfunc}
x_\pm &=&\frac{s+m_1^2-m_2^2}{2s}\pm \frac{q(s)}{\sqrt{s}} \,, \nonumber\\
G(s)^{\rm DR} &=& \frac{1}{16\pi^2}\left[ a(\mu^2) + \log\frac{m_2^2}{\mu^2}-x_+\log\frac{x_+-1}{x_+}
-x_-\log\frac{x_--1}{x_-} \right], 
\end{eqnarray}
being $\mu$ the regularization scale. In the dispersive representation of the function $G(s)$, $a(\mu^2)$ in the above equation corresponds to the subtraction constant, which has to be determined by the experimental or lattice inputs. There are also other ways to express the function $G(s)$, such as to include the form factors  or to introduce sharp momentum cutoff to regularize the divergent integral in Eq.~\eqref{eq.defg}~\cite{Oller:1997ti,Albaladejo:2016lbb}. The merit of the representation of the $G(s)$ function in Eq.~\eqref{eq.gfunc} is that it has the correct analytical behaviors. In the multiple-channel scattering case, $T(s)$ and $N(s)$ in Eq.~\eqref{eq.defut} become matrices spanned in the channel space and $G(s)$ corresponds to the diagonal matrix with non-vanishing elements given by Eq.~\eqref{eq.defg}. 

In order to study the lattice discrete energy levels, one has to introduce the finite-volume effects. This can be done by calculating the function $G(s)$ of Eq.~\eqref{eq.defg} in the finite cubic box of length $L$ with periodic boundary conditions. In the center of mass (CM) frame, the resulting finite-volume correction $\Delta G$ to the function $G(s)$ in the infinite volume reads~\cite{Doring:2011vk}  
\begin{eqnarray}\label{eq.deltag}
\Delta G &=& \frac{1}{L^3} \sum_{\vec{n}}^{|\vec{q}|<q_{\rm max}} I(|\vec{q}|) -   \int^{|\vec{q}|<q_{\rm max}} 
\frac{{\rm d}^3 \vec{q}}{(2\pi)^3} I(|\vec{q}|)\,, 
\end{eqnarray}
with 
\begin{eqnarray}
&& \vec{q}= \frac{2\pi}{L} \vec{n},\,(\vec{n} \in \mathbb{Z}^3) \,, \,
 I(|\vec{q}|) = \frac{w_1+w_2}{2w_1 w_2 \,[s-(w_1+w_2)^2]}\,, \,
w_i =\sqrt{|\vec{q}|^2+m_i^2}  \,. \nonumber 
\end{eqnarray} 
The three-momentum cutoff $q_{\rm max}$ is introduced to regularize separately the two terms in Eq.~\eqref{eq.deltag} and it is demonstrated that $\Delta G$ is rather weakly dependent on $q_{\rm max}$~\cite{Guo:2016zep}, due to the cancellation of its dependences of the two terms in that equation. The function $G(s)$ used in the finite-volume study is then given by  
\begin{eqnarray}\label{eq.gfuncfvdr}
\widetilde{G}^{\rm DR}= G^{\rm DR} + \Delta G \,.
\end{eqnarray}
The finite-volume spectra correspond to the solutions 
\begin{eqnarray}\label{eq.detfv}
 \det{[1 + N(s) \cdot \widetilde{G}^{\rm DR}]} =0\,.
\end{eqnarray}

To proceed the study of the $\jpsi\pi$ and $D\dvb$ coupled-channel scattering, labeled as 1 and 2 in order, one needs to parameterize the matrix elements of the $N(s)$ in Eq.~\eqref{eq.defut}. The $\jpsi\pi\to\jpsi\pi$ amplitude has been found to be tiny~\cite{Liu:2012dv} and we will simply fix the matrix element $N_{11}=0$. The transition amplitude $N_{12}$ for the $\jpsi\pi\to D\dvb$ will be approximated as a constant. According to the discussion in Ref.~\cite{Albaladejo:2016lbb} and references therein, there is no resonance if one also takes the constant approximation for $N_{22}$. At least one next-to-leading order term in the momentum expansion should be kept in order to generate the resonance in the coupled-channel scattering. Therefore we will take one non-vanishing energy dependent term in the $N_{22}$. The explicit expression for the $N(s)$ is given by 
\begin{eqnarray}\label{eq.defn}
N(s) = \bma
      N_{11}  & N_{12}  \\
      N_{12} & N_{22} \\
      \ema
 =   \bma
     0     \qquad &  c_{12}  \\
     c_{12} \qquad & c_{22}^{0} + \frac{c_{22}^{1}}{m_D} (\sqrt{s}-m_D-m_D^*)\\
   \ema\,.
\end{eqnarray}

\section{Lattice finite-volume spectra and relevant resonances}

Several lattice finite-volume energy levels from the $\jpsi\pi$ and $D\dvb$ scattering are recently calculated in Refs.~\cite{Cheung:2017tnt,Chen:2019iux}. Both studies in the two former references claim that a standard narrow resonance state around the $D\dvb$ threshold is disfavored, although more precise energy levels with different volumes and lighter pion masses are needed to reach definite conclusions.

\begin{figure}
\begin{center}
\includegraphics[width=0.7\textwidth]{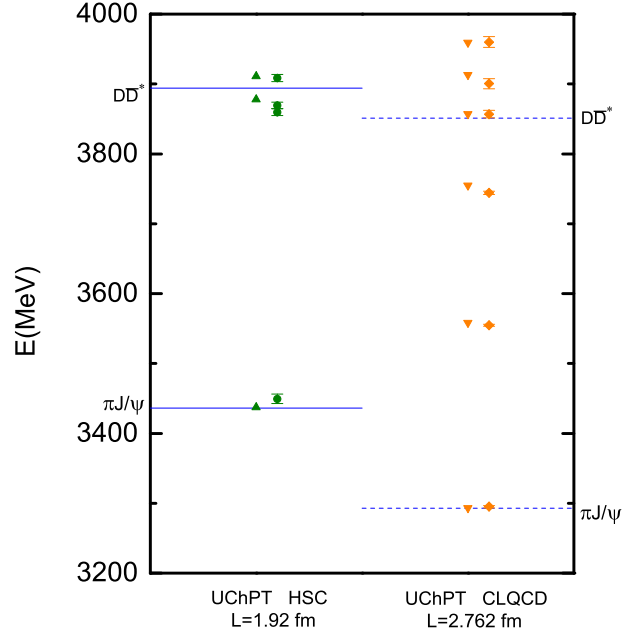}
\end{center}
\caption{The comparisons of our predictions and the lattice energy levels from Ref.~\cite{Cheung:2017tnt} (HSC) and Ref.~\cite{Chen:2019iux} (CLQCD). Notice that different meson masses are used in the two lattice simulations.}
\label{fig.fig1}
\end{figure}

Here we take an exploratory study to use the formalism in Eq.~\eqref{eq.detfv} to confront the lattice spectra in Refs.~\cite{Cheung:2017tnt,Chen:2019iux}. To fix the unknown parameters $c_{12}, c_{22}^{0}, c_{22}^{1}$ in Eq.~\eqref{eq.defn} and the subtraction constants in Eq.~\eqref{eq.gfunc}, which are assumed to be independent on the quark masses, we will use the $D\dvb$ amplitudes from an upcoming $\jpsi\pi$ and $D\dvb$ coupled-channel study in Ref.~\cite{guoprepare} to constrain the free parameters. We leave the details of the coupled-channel analyses in a future work~\cite{guoprepare}. Our primary aim here is to check whether it is possible to find a solution to reasonably describe the lattice spectra in Refs.~\cite{Cheung:2017tnt,Chen:2019iux}, while in the meantime there is a $Z_c$ resonant state. 

The comparisons of our predictions and the lattice spectra are given in Fig.~\ref{fig.fig1}. It is clear that our theoretical predictions are compatible with the lattice results within uncertainties. With the same parameters that give the energy levels in Fig.~\ref{fig.fig1}, we do find in the complex energy plane the $Z_c$ resonance poles, among which the most relevant one is $(3887.2 - i 1.5)$~MeV in the second Riemann sheet, although other shadow poles are also found in different Riemann sheets. However it is pointed out that none of the $Z_c$ poles correspond to the conventional narrow-width resonance, since they do not appear in the proper Riemann sheets and the phase shifts around the $D\dvb$ threshold do not pass ninety degrees as the conventional resonance. 

\section{Acknowledgments}
This work is partially supported by the NSFC under Grants No.~11975090,  No.~11575052, and the Natural Science Foundation of Hebei Province under Contract No.~A2015205205.

\end{document}